\documentclass[12pt]{article}
\input epsf.sty
\def\be{\begin{equation}} \def\ee{\end{equation}}
\def\bea{\begin{eqnarray}} \def\eea{\end{eqnarray}} \def\ba{\begin{array}}
\def\ea{\end{array}} \def\ben{\begin{enumerate}} \def\een{\end{enumerate}}

\newcommand{\eqn}[1]{(\ref{#1})}

\def\F{\Phi}

\def\br{\nonumber\\}

\def\dA'{\dot A}\def\dB'{\dot B}
\def\dA''{\ddot A}\def\dB''{\ddot B}
\begin{document}
{}~

\vskip .5cm
\centerline{\Large \bf
Alpha-prime corrections to space-like branes
}

\centerline{\Large \bf}

\vspace*{.5cm}

\centerline{  Shibaji Roy\footnote{E-mail: shibaji.roy@saha.ac.in}
 and Harvendra Singh\footnote{E-mail: h.singh@saha.ac.in}}

\vspace*{.25cm}

\centerline{ \it Saha Institute of Nuclear Physics} 
\centerline{ \it  1/AF Bidhannagar, Kolkata 700064, India}

\vspace*{.5cm}


\vskip2cm
\centerline{\bf Abstract} \bigskip

Space-like branes or S-branes are certain class of time dependent
solutions of M/string theories having isometries ISO($p+1$) $\times$
SO($d-p-2,1$), where $d=11,10$ respectively and have singularities
at $t=0$. In \cite{Roy:2006du}, we found that the asymptotically flat 
S-branes have the structure of generalized Kasner geometry near $t=0$. 
In this work we evaluate higher order $\alpha'$ 
corrections perturbatively 
to the heterotic string Kasner backgrounds to probe the singularity
at $t=0$. 
We generally find that the 
perturbative corrections do not permit us to reach the singular
point, as the supergravity framework fails near $t \sim {\sqrt \alpha'}$
blurring the 
origin of space-like singularities. This is analogous to the concept 
of stretched horizons in the case of black holes.

 \vfill \eject

\baselineskip=16.2pt

\section{Introduction}
M/String theories are known to admit a variety of time dependent solutions.
Space-like or S-branes \cite{Gutperle:2002ai} are certain class of such 
solutions possessing
an isometry ISO($p+1$) $\times$ SO($9-p,1$) for M-theory and ISO($p+1$) 
$\times$ SO($8-p,1$) for string theories. They can be
asymptotically non-flat \cite{Chen:2002yq} as well as flat 
\cite{Kruczenski:2002ap,Bhattacharya:2003sh}\footnote{The relations 
between these two classes of S-brane
solutions have been discussed in \cite{Roy:2002ik}.}. The 
asymptotically flat S$p$-branes
were constructed in ref.\cite{Bhattacharya:2003sh} and were found
\cite{Roy:2006du} to have a space-like or
cosmological singularity at $t=0$. As time dependent solutions, S-branes are
particularly interesting as they might help us to understand the time
dependent processes in string theory like brane-antibrane or non-BPS brane 
decay through rolling tachyon \cite{Sen:2002nu,Sen:2002in,Sen:2002an}. 
They are also interesting from the cosmological 
point of view and in particular they are known to give rise to four-dimensional
accelerating cosmologies 
\cite{Townsend:2003fx,Ohta:2003pu,Roy:2003nd,Ohta:2003ie,Roy:2006du}. But the 
acceleration in this case is transient, 
i.e., it occurs only for a finite interval 
of time. This feature persists for both asymptotically non-flat and flat
solutions \cite{Roy:2006du}. But, because of its transient nature, 
it does not give
rise to high enough e-folding necessary for realistic cosmology. However,
the inclusion
of higher curvature terms improves the situation and some such studies
have been reported in \cite{Maeda:2004hu,Akune:2006dg} and 
more recently in \cite{Niz:2007gq}.

In this paper we will be interested in another aspect of the asymptotically
flat S-brane solutions, namely, the cosmological singularity and see
the effect of stringy corrections to it.
In fact the recent interest in perturbative corrections to string theory has
revealed many  new insights in the understanding of various aspects of 
black hole singularities 
\cite{Dabholkar:2004yr,Dabholkar:2004dq,Sen:2004dp,Sen:2005pu,Sen:2005wa}. 
The leading world sheet corrections in 
some supersymmetric black hole geometries, which violate cosmic censorship
and have degenerate horizons with vanishing areas, in fact 
become finite when $\alpha'$ corrections are included. It has been the 
belief in string theory that generally the space and time singularities 
must get smoothed out when relevant stringy corrections are taken into account. 
This has partly been tested in case of (small) black holes
where quantum corrections are under control due to the large amount of 
supersymmetry. However, in  non-supersymmetric cases it is rather tricky, as 
the quantum corrections can be very large and even uncontrolled. But 
recent study in the case of extremal but non-supersymmetric black holes 
have led to not very different physics than in the case of 
supersymmetric ones \cite{Dabholkar:2006tb, Astefanesei:2006sy}. 

As we mentioned
our main interest in this work is to 
study the stringy corrections to the S-brane background particularly near
the singular point. But we should keep in mind that the time dependent 
backgrounds are 
non-supersymmetric and so the question of perturbative corrections is rather 
convoluted.   
The asymptotically flat S-branes were studied previously in \cite{Roy:2006du} 
and this has provided some interesting new insights, particularly, 
we found that the near $t=0$ limit 
of the S-brane solutions leads to generic Kasner type geometries
\cite{Das:2006dz}. Like
S-branes, Kasner geometries are inherently singular at $t=0$ and 
therefore are interesting cosmological solutions. Also they are much
simpler than full S-brane solutions and make a good 
example for the theoretical understanding of space-like singularities. 
It would be 
interesting to study the stringy corrections to these Kasner 
geometries\footnote{Stringy corrections to the bosonic string Kasner 
backgrounds were studied in \cite{Exirifard:2004ey}. We thank Qasem
Exirifird for informing us about this.} and see what 
effect they have on the space-like singularity
of this background.

The paper is organised as follows. In section-2 we review the $t\to 0$ 
limit of S-brane solutions and the resultant generalised Kasner 
cosmologies. We present the special case of S2-branes. In section-3 we 
obtain the $\alpha'$ corrections to the Kasner backgrounds in a heterotic 
string set up. We will consider two cases, namely, in the first case
the lowest order dilaton will be taken to be constant while in the
second case it will be non-constant.  The results are 
summarised in section-4.
  
\section{Generalized Kasner backgrounds}

In \cite{Roy:2006du} the generalized Kasner backgrounds were 
obtained as the near
`horizon' or $t \to 0$ limit of the S-brane solutions of type II string 
theory in the lowest order (without $\alpha'$ correction). The ten 
dimensional action we considered was,
\be\label{act}
S=\frac{1}{2\kappa^2}\int d^{10}x \sqrt{- g} \bigg[ R 
- \frac{1}{2} (\partial\phi)^2 - \frac{e^{a\phi}}{2\cdot n!} F_{[n]}^2\bigg] 
\ee
where $\kappa$ is related to the ten dimensional Newton's constant, 
$g = {\rm det}\,
(g_{\mu\nu})$, with $g_{\mu\nu}$ being the Einstein-frame metric, $R$
is the scalar curvature, $\phi$ is the dilaton and $F_{[n]}$ is the
$n$-form field strength. Also $a$ is the dilaton coupling given by
$a^2 = 4 - 6(n-1)/(n+2)$ for maximal supergravities. Depending on the
value of $n$ we can obtain various S$p$-brane (where $p=8-n$) solution 
from the
equations of motion following from \eqn{act} with the
appropriate ansatz for the metric and the form-field. We, however,
will be looking at S2-brane solution and so we will put $n=6$. The
reason for this is that the S2-brane has 3-dimensional Euclidean world-volume
and so, including the time direction we get the 4-dimensional world
whose cosmology we are interested in. Therefore, when dilaton is non-constant
$a=-1/2$ and when dilaton is constant, $a$ will be put to zero. Note that
$a=0$ does not correspond to maximal supergravity as we mentioned earlier.
The S2-brane solution of the above action has the form 
\cite{Bhattacharya:2003sh,Roy:2006du},
\bea\label{S2}
ds^2 &=& F^{\frac{12}{5\chi}}g^{\frac{1}{5}}\left(-\frac{dt^2}{g} + t^2 dH_6^2
\right) + F^{-\frac{4}{\chi}}\sum_{i=1}^3 (dx^i)^2\br
e^{2\phi} &=& F^{-\frac{32}{5\chi}a} g^{\delta}, \qquad 
F_{[6]}\,\,\,=\,\,\, b {\rm Vol}
(H_6)\eea
where 
\bea\label{functions}
F &=& g^{\alpha/2} \cos^2\theta + g^{-\beta/2} \sin^2\theta\br
g &=& 1 + 4\omega^5/t^5
\eea
In the above, $\alpha$, $\beta$, $\theta$, $\omega$ and $\delta$ are 
integration constants and $b$ is the charge parameter. Also $H_6$ is the
6-dimensional hyperbolic space and $dH_6^2$ is its line element while 
Vol($H_6$) is its volume form. Also in the above $\chi=6+ 8a^2/5$ and so,
$\chi= 32/5$ when $a=-1/2$ i.e. the dilaton is non-constant
and $\chi=6$ when dilaton is constant.
The paramaters in the solution just mentioned
are related as,
\bea\label{parameter}
b &=& \sqrt{\frac{160}{\chi}} (\alpha+\beta) \omega^5 \sin2\theta\br
\alpha &=& \pm \sqrt{\frac{6\chi - 15 \delta^2}{16}} + \frac{a\delta}{2}\br
\beta &=& \pm \sqrt{\frac{6\chi - 15 \delta^2}{16}} - \frac{a\delta}{2}
\eea
Now it is easy to check from \eqn{functions} that at early time i.e.
as $t\to 0$, the function $F(t)$ behaves as $F(t) \sim t^{-5\alpha/2}$.
Note that both the upper and the lower signs of $\alpha$ and $\beta$
gives exactly same behavior of $F(t)$. So, after some rescaling of 
coordinates and
redefining the time coordinate by $t^{(-3\alpha/\chi)+2} dt \to dt$ we 
can rewrite the configuration \eqn{S2} as follows,
\bea\label{genkasner}
ds^2 &=& -dt^2 + t^{-\frac{10\alpha}{3(\alpha-\chi)}} \sum_{i=1}^3 (dx^i)^2
+ t^{\frac{(6\alpha-\chi)}{3(\alpha-\chi)}}dR_6^2\br
e^{2\phi} & \sim & t^{\frac{-16a\alpha+ 5 \delta \chi}{3(\alpha-\chi)}}  \qquad 
F_{[6]}\,\,\,=\,\,\,0
\eea
where `$\sim $' means upto a constant which can be absorbed by
a constant shift of the dilaton. 
Note that while obtaining  eq.\eqn{genkasner} from eq.\eqn{S2} we had to 
replace
$dH_6^2$ by $dR_6^2$, i.e. the flat space and this is because near $t=0$,
the overall radius of the hyperbolic space becomes very large as can be 
seen from
\eqn{S2} and so effectively this space becomes flat. Also, because of 
that reason the charge parameter $b$ must vanish identically which implies
that in that limit the parameter $\theta$ must vanish. The parameter 
$\omega$ can be absorbed into the coordinate rescaling and the redefinition
of the dilaton field. So, out of the three independent parameters $\omega$, 
$\theta$ and $\delta$, the solution near $t=0$ is characterized by a single
parameter $\delta$ (or $\alpha$ which is related to $\delta$ by 
\eqn{parameter}). By defining the various exponents of $t$ appearing in
the metric and the dilaton in \eqn{genkasner} as
\bea\label{gkasner}
ds^2 &=& -dt^2 + t^{2p} \sum_{i=1}^3 (dx^i)^2
+ t^{2q}dR_6^2\br
e^{2\phi} &\sim & t^{2\gamma}
\eea
where
\be\label{pqgamma}
p=-\frac{5\alpha}{3(\alpha-\chi)},\qquad q=\frac{6\alpha-\chi}{6(\alpha-\chi)},
\qquad \gamma=\frac{-16a\alpha+5\delta\chi}{6(\alpha - \chi)}
\ee
we find that they satisfy,
\be\label{relation}
3p+6q=1, \qquad {\rm and} \qquad 3p^2+6q^2=1-\frac{1}{2}\gamma^2
\ee
This is precisely the generalized Kasner geometry \cite{Das:2006dz}
one obtains as the near 
`horizon' or near $t=0$ limit of the asymptotically flat S2-brane solution of
type II string theory. We remark that in the above $a$ can take only two 
values, namely, $a=0$ (when dilaton is constant in the lowest order) and 
$a=-1/2$ (when dilaton is
non-constant in the lowest order). Consequently, $\chi$ can also take 
two values, namely, $\chi=6$
(when dilaton is constant) and $\chi=32/5$ (when dilaton is non-constant). 
Therefore, when dilaton is constant in the lowest order  
we have $\alpha=\pm 3/2$ and $\gamma=0$.
Now since there is no dilaton field the background in this case reduces to 
Kasner geometry. Putting these values in \eqn{pqgamma} we find that $p$
and $q$ take the following two sets of values,
\bea\label{pq}
&& (i)\,\, p = 5/9, \qquad q = -1/9\br
&& (ii)\,\, p = -1/3, \qquad q = 1/3
\eea
We will study the stringy corrections to these backgrounds in the next section.
Next, we discuss the case when dilaton is non-constant at the lowest order. 
In this case as we
mentioned  
$a=-1/2$ and $\chi=32/5$. Now since in this case we have two relations
\eqn{relation} with three unknowns $p$, $q$ and $\gamma$, there will be
infinite number of solutions. We will discuss two simple cases and their
stringy corrections in the next section. The first case is
\be\label{1st}
(a)\,\, p=q=1/9, \qquad {\rm and} \qquad \gamma=-4/3
\ee
and the second case is
\be\label{2nd}
(b)\,\, p=\frac{7 - 8\sqrt{3}}{39}, \qquad q=\frac{3 +
4\sqrt{3}}{39} \qquad {\rm and} \qquad \gamma = -4\left(\frac{3+ 4\sqrt{3}}
{39}\right)
\ee
These are the form of the metric and the dilaton in Einstein frame. But since
we have a non-trivial dilaton in the lowest order, the form of the metric 
will be different in
the string frame. We will give the form of the metric in the string frame. The
reason behind this is that the $\alpha'$ correction terms are explicitly known
\cite{Freeman:1986zh,Gross:1986mw,Hull:1987yi}
in the string frame. If we write the string frame metric and the dilaton as
\bea\label{string}
ds^2 &=& -dt^2 + t^{2p'}\sum_{i=1}^3 (dx^i)^2 + t^{2q'} dR_6^2\br
e^{2\phi} &\sim & t^{2\gamma'}
\eea
then the two solutions given above in $(a)$ and $(b)$ take the form in string
frame as,
\bea\label{2ndstring}
&& (i')\,\, p'=q'=-1/3, \qquad {\rm and} \qquad \gamma'= -2\br
&& (ii')\,\, p'= -\frac{1}{\sqrt{3}}, \qquad q'=0, \qquad {\rm and} \qquad
\gamma' = -\frac{1}{2}\left(1+\frac{1}{\sqrt 3}\right) 
\eea
We will study the stringy corrections to these backgrounds in the next section.

\section{Alpha-prime corrections}

Although we had discussed S-branes in type II theories in the previous section,
here we will restrict ourselves to the special case of heterotic string
background. Since the S2-brane we described in \eqn{genkasner}
is chargeless, it is
also a solution to the heterotic string theory. The reason for this restriction
is that it is the heterotic string theory which contains a non-trivial 
$\alpha'$ correction term in the form of Gauss-Bonnet term, whereas, in 
type II string theory the first non-trivial correction comes at 
$\alpha'^3$ order \cite{Freeman:1986zh,Gross:1986mw,Hull:1987yi}
and so the calculation becomes more involved. First we 
will discuss the case where the dilaton is constant in the lowest order.

\subsection{Constant lowest order dilaton and the cosmologies:}

We first investigate the special case of heterotic backgrounds 
where the 
dilaton is constant in the lowest order.
\be\label{eq1}
ds^2= -dt^2+ A(t)^2 \sum_{i=1}^3(dx^i)^2 + B(t)^2 \sum_{j=4}^9 (dx^j)^2
\ee
Note here that the metric has the isometry $R^+ \times SO(3) \times SO(6)$
and the time coordinate range $0 \leq t \leq \infty$.
The string coupling $g_s$ is 
taken to be  very small, so we shall neglect string loop 
corrections. 
One can take the $x^j$'s to be the coordinates on some compact 
Ricci-flat six-manifold but here we 
take it to be simple toroidal case. Thus in this solution dilaton 
does not vary at least in the lowest order, while tensor and gauge fields 
are switched off. We substitute the above ansatz 
in the action (which is basically the Einstein-Hilbert action since the 
dilaton is constant) and then minimise the resulting action with respect to $A$ 
and $B$.
The metric in \eqn{eq1} solves vacuum Einstein equations  
with
\bea\label{ab}
&&(i) ~ A(t)=t^{5/9}, ~~~~~ B(t)= t^{-1/9} \br
&&(ii) ~    A(t)=t^{-1/3}, ~~~~~ B(t)= t^{1/3} 
\eea
Note that this is precisely the solution we obtained in cases $(i)$ 
and $(ii)$ in \eqn{pq}. This is not surprising since \eqn{pq} also
represents the solutions to the vacuum Einstein equation in ten 
dimensions (note
that these solutions are obtained from \eqn{genkasner} when the dilaton is
constant and the form field is put to zero).
However, we note that for both the solutions there is an 
essential singularity at $t=0$, where Riemann tensor 
blows up.  The scalar curvature $R$ vanishes identically 
on-shell, but
the contractions like
 $$R_{MNPQ}R^{MNPQ}\sim {1\over t^4}$$  blow up at the singularity. 
Since the latter term appears as  the leading world sheet 
correction in the heterotic string theory it 
becomes imperative to include them in the string action. However, as
soon as we include the higher order $\alpha'$ correction terms it
is not true that the dilaton will still remain constant\footnote{By the
same token one might think that the NSNS field $H_{[3]}$ also can not remain
zero once the $\alpha'$ corrections are included. Indeed they can
be generated via the gravitational Chern-Simons term. But it can be easily
checked that for the Kasner backgrounds the Chern-Simons term is in fact
vanishing and so, we can consistently put it to zero.
We thank
Qasem Exirifard for very useful comments which made us rethink
on this.} \cite{Callan:1988hs}. So, we have
to consider the full heterotic string effective action including the
dilaton in the leading order of $\alpha'$ and is given by 
\cite{Gross:1986mw,Hull:1987yi}
\bea
S&=&{1\over 2\kappa^2}\int d^{10}x \sqrt{- g}\Phi \bigg[ R
+4 (\partial\phi)^2
\br
&&+{\alpha'\over 8} \left( R^2_{GB}-
16 (G^{MN} \partial_M\phi  \partial_N\phi -\nabla^2\phi  (\partial\phi)^2
+ ( \partial\phi)^4) \right )
\bigg] \label{action}
\eea
Here $R_{GB}^2 = R_{MNPQ} R^{MNPQ} - 4 R_{MN} R^{MN} + R^2$ is the 
Gauss-Bonnet term, $\Phi \equiv e^{-2\phi}$ is defined for simplicity
and $G_{MN} = R_{MN} - g_{MN} (R/2)$ is the Einstein's tensor. Note that
\eqn{ab} is still a solution to the above action \eqn{action} in the leading
order.
Now the various field equations with $\alpha'$ terms 
will be solved 
order by order in the neighborhood of these leading order solutions $(i)$ and 
$(ii)$. Let us plug the ansatz \eqn{eq1}  in the above 
action \eqn{action} to get,
\bea \label{actionsimple}
S&\sim&\int dt \bigg[ 6\Phi \left( A B^6 \dot A ^2 + 5 A^3 B^4
\dot B^2 + 6 B^5 A^2 \dot B\dot A\right) + 6\dot \Phi\left(2 A^3B^5
\dot B + A^2 B^6
\dot
A\right) +  \Phi A^3 B^6 \frac{\dot\Phi^2}{\Phi^2} \bigg]  \br
&& +\alpha'\left[ \Phi \left(15 B^2A^3 \dot B^4 +60 B^3 A^2 \dot B^3 \dot A
+45 B^4 A
\dot B^2\dot A^2 + 6 B^5 \dot B\dot A^3\right)\right.\br
&& \qquad +\dot\Phi \left( 20 B^3 A^3
\dot B^3 +
B^6 \dot A^3 +45 B^4 A^2 \dot A \dot B^2 +18 B^5 A \dot A^2\dot B\right) \br
&& \qquad + \frac{1}{2}\Phi \left\{\left(15 A^3 B^4 {\dot B}^2
+ 18 A^2
B^5 {\dot A}{\dot B} + 3 A B^6 {\dot A}^2\right)\frac{{\dot \Phi}^2}{\Phi^2}
\right.\br
&&\qquad\qquad -\left.\left. \frac{1}{2}\left(A^2B^6{\dot A} + 2 A^3 B^5
{\dot B}\right)\frac{
{\dot \Phi}^3}{\Phi^3} + \frac{1}{12} A^3 B^6 \frac{{\dot \Phi}^4}{\Phi^4}
\right\}\right]
\eea
Note that the total derivative terms in \eqn{actionsimple} 
have been dropped since they do not contribute to the
equations of motion.  
It can be noted that leading order terms are quadrartic in
 time derivatives  
while next order terms have up to four derivatives in them. 
The equations of motion for the various fields can be 
straightforwardly obtained from the variation of the above action with
respect to $A(t)$, $B(t)$ and $\Phi(t)$ and they are given respectively as,
\bea\label{Aeqn} 
&&\frac{3 B^4}{\F} \left(30 \F^2 A^2 {\dot B}^2+12 \F B A 
\left(2 \F {\dot B} {\dot A}+A \left(\dot\Phi {\dot B}+\F {\ddot B}\right)
\right)+
B^2
\left(2 \Phi^2 {\dot A}^2-A^2 \left({\dot\Phi}^2-2 \F \ddot\Phi\right)
\right.\right.\br
&&\left.\left.+4 \F A 
\left(\dot\Phi \dot A+\F \ddot A\right)\right)\right)+ \alpha'
\bigg[ \frac{3 B^2}{8 \Phi^3}
\left(360 \Phi^4 A^2 {\dot B}^4+480 \Phi^3 B A {\dot B}^2 
\left(2 \F \dot B \dot A\right.\right.\br
&& \left.+A \left(\dot\Phi \dot B+\F \ddot B\right)\right)
+B^4 \left(A^2
\left({\dot\Phi}^4-2 \F {\dot\Phi}^2 \ddot\Phi\right)+8 \F A \dot\Phi 
\left(-{\dot\Phi}^2 \dot A+2 \F \dot A \ddot\Phi+\F \dot\Phi \ddot A\right)
\right.\br
&& \left.+4 \Phi^2 \dot A \left({\dot\Phi}^2
\dot A+2 \F \dot A \ddot\Phi+4 \F \dot\Phi \ddot A\right)\right)
+60 \Phi^2 B^2 \dot B \left(6 \Phi^2 \dot B {\dot A}^2+A^2 
\left({\dot\Phi}^2 \dot B+2 \F
\dot B \ddot\Phi\right.\right.\br
&& \left.\left. +4 \F \dot\Phi \ddot B\right)+4 \F A 
\left(3 \dot\Phi \dot B \dot A+2 \F \dot A \ddot B+\F \dot B \ddot A\right)
\right)+24 \F B^3
\left(A^2 \dot\Phi \left(-{\dot\Phi}^2 \dot B+2 \F \dot B \ddot\Phi+\F 
\dot\Phi \ddot B
\right)\right.\br
&&+2 \F A \left({\dot\Phi}^2 \dot B \dot A+2 \F \dot B \dot A 
\ddot\Phi+2
\F \dot\Phi \left(\dot A \ddot B+\dot B \ddot A\right)\right)\br
&& \left.\left.+2 \Phi^2 \dot A 
\left(3 \dot\Phi \dot B \dot A+\F \left(\dot A \ddot B+2 \dot B \ddot A
\right)\right)
\right)\right)\bigg] = 0
\eea

\bea\label{Beqn}
&&\frac{6 B^3 A}{\Phi} \left(20 \Phi^2 A^2 {\dot B}^2+10 \Phi 
B A \left(3 \Phi \dot B \dot A+A \left(\dot\Phi \dot B+\Phi 
\ddot B\right)\right)+B^2
\left(6 \Phi^2 {\dot A}^2-A^2 \left({\dot\Phi}^2-2 \Phi \ddot\Phi\right)\right.
\right.\br
&&\left.\left.+6 \Phi A \left(\dot\Phi \dot A+\Phi \ddot A
\right)\right)\right)
+ \alpha'\bigg[ \frac{3 B}{4\Phi^3}
\left(120 \Phi^4 A^3 {\dot B}^4+240 \Phi^3 B A^2 {\dot B}^2 
\left(3 \Phi \dot B \dot A\right.\right.\br
&&\left.+A \left(\dot\Phi \dot B+\Phi \ddot B\right)\right)
+B^4 \left(A^3
\left({\dot\Phi}^4-2 \Phi {\dot\Phi}^2 \ddot\Phi\right)+24 \Phi^3 {\dot A}^2 
\left(\dot\Phi \dot A+\Phi \ddot A\right)+12 \Phi A^2 \dot\Phi 
\left(-{\dot\Phi}^2 \dot A\right.\right.\br
&&\left.\left.+2 \Phi
\dot A \ddot\Phi+\Phi \dot\Phi \ddot A\right)+12 \Phi^2 A \dot A 
\left({\dot\Phi}^2 \dot A+2 \Phi \dot A \ddot\Phi+4 \Phi \dot\Phi 
\ddot A\right)\right) +40 \Phi^2 B^2
A \dot B \left(18\Phi^2 \dot B {\dot A}^2\right.\br
&&\left.+A^2 \left({\dot\Phi}^2 \dot B
+2 \Phi \dot B \ddot\Phi+4 \Phi \dot\Phi \ddot B\right)+6 \Phi A 
\left(3 \dot\Phi \dot B
\dot A+2 \Phi \dot A \ddot B+\Phi \dot B \ddot A\right)\right)
+20 \Phi B^3 \left(6 \Phi^3 \dot B {\dot A}^3\right.\br
&&+A^3 \dot\Phi 
\left(-{\dot\Phi}^2 \dot B+2 \Phi \dot B
\ddot\Phi+\Phi \dot\Phi \ddot B\right)+3 \Phi A^2 \left({\dot\Phi}^2 
\dot B \dot A
+2 \Phi \dot B \dot A \ddot\Phi+2 \Phi \dot\Phi \left(\dot A 
\ddot B+\dot B \ddot A
\right)\right)\br
&& \left.\left.+6\Phi^2 A \dot A \left(3 \dot\Phi \dot B \dot A+\Phi 
\left(\dot A \ddot B+2 \dot B \ddot A\right)\right)\right)\right) \bigg]
= 0
\eea

\bea\label{Feqn}
&&\frac{B^4 A}{\Phi^2} \left(30 \Phi^2 A^2 {\dot B}^2+12 \Phi B 
A \left(3 \Phi \dot B \dot A+A \left(\dot\Phi \dot B+\Phi \ddot B
\right)\right)+B^2
\left(6 \Phi^2 {\dot A}^2-A^2 \left({\dot\Phi}^2-2 \Phi \ddot\Phi\right)
\right.\right.\br
&& \left.\left.+6 \Phi A \left(\dot\Phi \dot A+\dot\Phi \ddot A\right)
\right)\right)
+\alpha'\bigg[ \frac{B^2}{8\Phi^4}
\left(360 \Phi^4 A^3 {\dot B}^4+480 \Phi^3 B A^2 {\dot B}^2 
\left(3 \Phi \dot B \dot A+A \left(\dot\Phi \dot B\right.\right.\right.\br
&&\left.\left.+\Phi \ddot B\right)
\right)
 +B^4
\left(A^3 \left(-3 {\dot\Phi}^4+4 \Phi {\dot\Phi}^2 \ddot\Phi\right)+
24 \Phi^3 {\dot A}^2 
\left(\dot\Phi \dot A+\Phi \ddot A\right)-6 
\Phi A^2 \dot\Phi \left(-2 {\dot\Phi}^2
\dot A+2 \Phi \dot A \ddot\Phi\right.\right.\br
&&\left.\left.+\Phi \dot\Phi \ddot A\right)
+24 \Phi^2 A \dot A 
\left(-{\dot\Phi}^2 \dot A+\Phi \dot A \ddot\Phi+2 \Phi \dot\Phi \ddot A
\right)\right)+120
\Phi^2 B^2 A \dot B \left(9 \Phi^2 \dot B {\dot A}^2+A^2 \left(-{\dot\Phi}^2 
\dot B\right.\right.\br
&&\left.\left. +\Phi \dot B \ddot\Phi+2 \Phi \dot\Phi \ddot B\right)+3 \Phi A 
\left(3
\dot\Phi \dot B \dot A+2 \Phi \dot A \ddot B+\Phi \dot B \ddot A\right)\right)
+12 \Phi B^3 \left(12 \Phi^3 \dot B {\dot A}^3\right.\br
&&+A^3 \dot\Phi \left(2 
{\dot\Phi}^2
\dot B-2 \Phi \dot B \ddot\Phi-\Phi \dot\Phi \ddot B\right)+12 \Phi A^2 
\left(-{\dot\Phi}^2 \dot B \dot A+\Phi \dot B \dot A \ddot\Phi+\Phi \dot\Phi
\left(\dot A \ddot B+\dot B
\ddot A\right)\right)\br
&&\left.\left.+12 \Phi^2 A \dot A \left(3 \dot\Phi \dot B \dot A
+\Phi \left(\dot A \ddot B+2 \dot B \ddot A\right)\right)\right)\right) 
\bigg]
= 0
\eea 
The corrected solutions are obtained by taking the perturbative ansatz
\bea\label{sol2}
A(t)&=&A_0(t) +\alpha' A_1(t)+O(\alpha'^2),\br
B(t)&=&B_0(t) +\alpha' B_1(t)+O(\alpha'^2),\br
\Phi(t) &=& \Phi_0(t) + \alpha' \Phi_1(t) + O(\alpha'^2) 
\eea 
where we  take $A_0$, $B_0$ and $\Phi_0$ as the known lowest order solutions.
Note that in this case the zeroth order dilaton is constant and so, 
$\Phi_0 = $ constant. 
We remark that the perturbative ansatz are good only in the time region
where $\alpha' A_1(t)\ll A_0(t)$ , $\alpha' B_1(t)\ll B_0(t)$ and
$\alpha' \Phi_1(t) \ll \Phi_0(t)$. The 
moment these bounds are violated we will be making wrong move. As we will
see that the perturbative ansatz are good particularly in the asymptotic 
regions.

By substituting \eqn{sol2} in \eqn{Aeqn}, \eqn{Beqn} and \eqn{Feqn} 
it is easy to see that to zeroth order in $\alpha'$, these equations 
take the forms
\bea \label{cdzero}
&& 15 A_0^2 {\dot B}_0^2 + 6 A_0 B_0\left(2{\dot A}_0 {\dot B}_0 + A_0
{\ddot B}_0\right) + {\dot A}_0^2 B_0^2 + 2 A_0 {\ddot A}_0B_0^2 = 0\br 
&& 10 A_0^2 {\dot B}_0^2 + 5 A_0 B_0\left(3{\dot A}_0 {\dot B}_0 + A_0
{\ddot B}_0\right) + 3{\dot A}_0^2 B_0^2 + 3 A_0 {\ddot A}_0 B_0^2 = 0\br
&& 5 A_0^2 {\dot B}_0^2 + 2 A_0 B_0\left(3{\dot A}_0 {\dot B}_0 + A_0
{\ddot B}_0\right) + {\dot A}_0^2 B_0^2 +  A_0 {\ddot A}_0B_0^2 = 0\br
\eea

\noindent{\bf Case(i)}:

Let us 
first take the case when
$A_0(t)=t^{5/9}$, $B_0(t)=t^{-1/9}$ and $\Phi_0(t) = \Phi_0 =$ constant as  
in the case $(i)$ which solve \eqn{cdzero}. Now substituting \eqn{sol2} with 
these values of $A_0$, $B_0$ and $\Phi_0$
in \eqn{Aeqn}, \eqn{Beqn} and \eqn{Feqn} we get three equations at the order 
$\alpha'$ involving $A_1(t)$, $B_1(t)$ and $\Phi_1(t)$ as follows,
\bea\label{a1b1f1}
&& \frac{5}{81}\frac{1}{t^{\frac{32}{9}}} + {1\over t^{{5\over 9}}}\left(
{4\over 3} {\dot {\Phi}}_1 + 6 {\dot B}_1 + 6 {\dot A}_1\right)
+ t^{{4\over 9}}\left(3 {\ddot \Phi}_1 + 18 {\ddot B}_1 + 6 {\ddot A}_1
\right) = 0\br
&& {35\over 81}{1\over t^{{26\over 9}}} + t^{{1\over 9}}\left(
{10\over 3} {\dot {\Phi}}_1 + 15 {\dot B}_1 + 15 {\dot A}_1\right)
+ t^{{10\over 9}}\left(3 {\ddot \Phi}_1 + 15 {\ddot B}_1 + 9 {\ddot A}_1
\right) = 0\br
&& {115\over 729}{1\over {t^3}} + \left(
{\dot {\Phi}}_1 + {16\over 3} {\dot B}_1 + {14\over 3} {\dot A}_1\right)
+ t\left( {\ddot \Phi}_1 + 6 {\ddot B}_1 + 3 {\ddot A}_1
\right) = 0\br
\eea
$A_1(t)$, $B_1(t)$ and $\Phi_1(t)$ can be solved from \eqn{a1b1f1} and we find, 
\bea 
A_1(t)&=& -{775\over 13122} t^{-{13\over9}},\br
B_1(t)&=& {10\over 6561} t^{-{19\over9}},\br
\Phi_1(t) &=& {1 \over g_s^2}{115\over 1458} t^{-2}
\eea
Note, as we mentioned earlier, that even though the lowest order $\Phi$ i.e.
$\Phi_0$ was kept constant, higher order correction makes the dilaton 
non-trivial.
Thus the complete solution up to  first order in $\alpha'$ is:
\be \label{sol3}
A(t)=t^{5\over 9}\left(1 -{775\over 13122} \frac{\alpha'}{t^2}\right),\quad
B(t)=t^{-{1\over 9}}\left(1+{10\over 6561} \frac{\alpha'}{t^2}\right),\quad
\Phi(t)={1\over g_s^2}\left(1+{115\over 1458} {\alpha' \over t^2}\right) 
\ee
where we have taken $\Phi_0 = {\rm constant} = 1/g_s^2 \gg 1$. Now
the metric and the dilaton becomes,
\bea\label{eq1a}
ds^2 &=& -dt^2+ t^{10\over 9}\left(1-{775\over 6561} {\alpha' \over t^2 
}\right)\sum_{i=1}^3 (dx^i)^2 
+ t^{-{2 \over 9}}\left(1+{20\over 6561} {\alpha' \over t^2}\right) 
\sum_{j=4}^9 (dx^j)^2\br
e^{2\phi} &=& g_s^2\left(1 - {115\over 1458} {\alpha'\over t^2}\right)
\eea
From the above 
we notice that the $\alpha'$ corrections are such that the metric 
components involving $A(t)$   
identically  vanish at some time $t=t_a$. The quantity $t_a$ can 
be obtained by taking 
$A(t)^2|_{t=t_a}=0$ in \eqn{eq1a}, we get
\be
t_a= \sqrt{775\over 6561}{\sqrt{\alpha'}} \ .
 \ee
While the determinant of the metric is
$${\rm Det}(-g)=t^2 \left(1-{245\over 
729}{\alpha'\over t^2}\right)+O(\alpha'^2) $$
which vanishes for $t=t_g$ where $$t_g= \sqrt{245\alpha'\over 729}.$$ Note 
that we have $t_g>t_a$. 
Thus the metric becomes degenerate at a finite time interval $t_g$ away 
from $t=0$.
The $t\gg t_g$ is the region where our perturbative calculation could be 
trusted. While for $t\le t_g$ the perturbative analysis will break 
down. Also, in other words, the physics becomes fuzzy 
from the supergravity point 
of view. However we still need to evaluate various curvature quantities 
to see if 
those remain finite at $t > t_g$. If that is the case we can say that 
$t \simeq t_g$ is some kind of a {\it `horizon of time'} behind that a 
cosmological 
singularity is hidden. Some of these quantities are listed below
\bea
&& R=-{230\over 243}{\alpha'\over t^4} +O(\alpha'^2)\br   
&& R_{MNPQ}R^{MNPQ}={1840\over 729 t^4} 
+O(\alpha')\br   
&& R_{MN}R^{MN}=0 +O(\alpha')\br   
&& R^2_{GB}={1840\over 729 t^4}+O(\alpha')  
\eea
By looking at the above expressions we find for $t> t_g$ all these 
quantities stay finite and
well defined. The actual singularity is  at $t=0$ where these quantities 
will blow up. But, {\it strictly speaking, these expressions are not valid 
 when 
$t \le t_g$.} and we cannot make any firm conclusion on the nature of the 
singularity. But for $t>t_g$ the spacetime makes perfect sense.

\noindent{\bf Case(ii)}:

Let us next look at the other solution when
$A_0(t)=t^{-1/3}$, $B_0(t)=t^{1/3}$ and $\Phi_0$ = constant as
in the case $(ii)$ which also solve \eqn{cdzero}. Again using \eqn{sol2} with
these values of $A_0$, $B_0$ and $\Phi_0$
in \eqn{Aeqn}, \eqn{Beqn} and \eqn{Feqn} we get three equations at the 
order $\alpha'$ involving $A_1(t)$, $B_1(t)$ and $\Phi_1(t)$ as follows,
\bea\label{a1b1f12}
&& \frac{5}{9}\frac{1}{t^{\frac{8}{3}}} + t^{{1\over 3}}\left(
4 {\dot {\Phi}}_1 + 30 {\dot B}_1 + 6 {\dot A}_1\right)
+ t^{{4\over 3}}\left(3 {\ddot \Phi}_1 + 18 {\ddot B}_1 + 6 {\ddot A}_1
\right) = 0\br
&& {7\over 9}{1\over t^{{10\over 3}}} + {1\over t^{1\over 3}}\left(
2{\dot {\Phi}}_1 + 15 {\dot B}_1 + 3 {\dot A}_1\right)
+ t^{{2\over 3}}\left(3 {\ddot \Phi}_1 + 15 {\ddot B}_1 + 9 {\ddot A}_1
\right) = 0\br
&& {1\over 3}{1\over {t^3}} + \left(
{\dot {\Phi}}_1 + 8 {\dot B}_1 + 2 {\dot A}_1\right)
+ t\left( {\ddot \Phi}_1 + 6 {\ddot B}_1 + 3 {\ddot A}_1
\right) = 0\br
\eea
The solutions to the above two equations \eqn{a1b1f12} are,
\be\label{a1b1sol}
A_1(t) = -\frac{1}{54} t^{-\frac{7}{3}}, \quad 
B_1(t) = -\frac{1}{27} t^{-\frac{5}{3}}, \quad
\Phi_1(t) = {1\over g_s^2}{1\over 6t^2}
\ee
So, the 
corrected solution upto first order in $\alpha'$ is
\be \label{sol5}
A(t)=t^{-{1\over 3}}\left(1-{1\over 54} \frac{\alpha'}{t^2}\right),~~~
B(t)=t^{1\over 3}\left(1-{1\over 27} \frac{\alpha'}{t^2}\right),~~~
\Phi(t)={1\over g_s^2} \left(1 + {\alpha' \over 6t^2}\right)
\ee
Here there appears to be more than one horizon. The outer most horizon is 
obtained by solving $B(t_b)^2=0$. It gives
\be
t_b= \sqrt{2\alpha'\over 27}
\ee
We obtain the inner horizon at $t=t_a$ where $A(t_a)^2=0$. But, 
$${\rm Det}(-g)=t^2 \left(1-{5\over 9}{\alpha'\over t^2}\right)
+O(\alpha'^2) $$ and the 
determinant becomes degenerate for $t=t_g = \sqrt{{5\alpha'\over 9}}$ and 
once again note that $t_g>t_b>t_a$.   

The curvature 
invariants are evaluated as
\bea
&& R=-2{\alpha'\over t^4} +O(\alpha'^2)\br   
&& R_{MNPQ}R^{MNPQ}={16\over 3 t^4} 
+O(\alpha')\br   
&& R_{MN}R^{MN}=0 
+O(\alpha')\br   
&& R^2_{GB}={16\over 3 t^4}+O(\alpha')\ .
\eea
With $\alpha'$ corrections included the backgrounds are no longer 
Ricci-flat. But
these quantites stay finite for $t> t_g$.
So it can be concluded
that higher derivative corrections for the simple Kasner geometries  
are very 
important in the neighborhood of $t=t_g$. First they resolve the 
cosmological
singularity by hiding the singularity 
behind the time horizon. It is much like as we find a stretched horizon at 
 string length away from the naked singularity for the 
BPS black holes which have vanishing horizon area classically. 
Had we got it differently, we would have been 
a bit disappointed. The conclusions are naive, that it is meaningless in 
the low energy 
supergravity set up to talk of the cosmological events (decay or 
creation) 
whose lifespan is smaller than string time 
$\sqrt{\alpha'}$. Second, in order to discuss or resolve space-like 
singularities, for example a Big-bang, 
we need to employ full string  theory.  

Specially, for both the pure Kasner 
geometries above we had string coupling fixed at very low value. But for 
other backgrounds the dilaton will generally vary in time and the string 
coupling may 
become very large at $t=0$.

\subsubsection{Spontaneous compactification}

It is worth noting that the $\alpha'$ corrected 
solutions for large $t$, i.e. when $t\gg \sqrt{\alpha'}$, do become usual 
Kasner solutions. 
The sizes and radii of the spatial directions change 
with time for the Kasner solutions. 
Particularly in the case $(i)$, for large $t$ the size 
of the six-dimensional 
space becomes naturally small and so it  can be compactified. We 
compactify it on  $T^6$ and  find that the 
corresponding 4-dimensional Einstein metric becomes
\be
ds^2= g_s^{-2}\left(-d\tau^2 + a(t)^2 \left((dx^1)^2+(dx^2)^2+(dx^3)^2
\right)\right) 
\ee
where $\tau={3\over2}t^{2\over3} (1- {15\over 6561}{\alpha'\over t^2})+ 
O(\alpha'^2)$ and $a(t)=t^{2\over 9} (1- {755\over 2\cdot 6561}{\alpha'\over 
t^2})$. The 4-dimesional dilaton is 
$$e^{2\phi_4}=g_s^2 t^{6\over 9}\left(1+{20\over 6561}{\alpha'\over t^2}
\right)^{-3}$$
plus there is volume modulus from compactification.
From this we 
determine that  the scale factor
\be
a(\tau)= \left({2\over3} \tau\right)^{1/3}\left(1-{745\over 388} 
{\alpha'\over \tau^3}\right) 
+ O(\alpha'^2)   .
\ee
So we can now evaluate the Hubble rate
\be\label{dfta}
H={1\over a}{d a \over d\tau}={1\over 3\tau}\left(1 + {745\over 432} 
{\alpha'\over\tau^3}\right) + O(\alpha'^2) .
\ee
and the deceleration rate 
\be\label{dft}
{1\over a}{d^2 a \over d\tau^2}=-\left({2\over 9} + {3725\over 1944} 
{\alpha'\over\tau^3}\right)\frac{1}{\tau^2} + O(\alpha'^2) .
\ee
Since we have horizon at $t=t_g=\sqrt{245\alpha'\over 729}$ which 
corresponds to $\tau_g^3\simeq 1.13 \alpha'$. 
As we see $\ddot a<0$, the cosmological expansion at late times, i.e.  
for $\tau^3 > \alpha'$, is always decelerating one. 
From the eq. \eqn{dfta} we see that the corrections tend to improve the
Hubble rate such that the quantity $\left({1\over 3} + {745\over 3 \cdot 432} 
{\alpha'\over\tau^3}\right)\sim .38$ when $\tau^3=11.3 \alpha'$. Also the  
deceleration rate is such that it initially decelerates faster than 
${2\over 9}$
if the corrections are included. In fact we can calculate the corrected
deceleration rate as $\left({2\over 9} + (1.92)(.089)\right) \sim 0.24$
which is very close to the value for radiation dominated phase of the
universe. It appears as if the universe was decelerating faster, as in
the radiation dominated phase, initially.

\subsection{ $O(\alpha')^2$ corrections}

It is straightforward to evaluate more higher order corrections to the 
solutions. Note that the heterotic action does not receive 
any $O(\alpha'^2)$ 
corrections for the Kasner background we have chosen \cite{Gross:1986mw}. 
Nevertheless, we can obtain 
the corrections in the solutions by solving 
equations to second order in $\alpha'$. So we take the ansatz
\bea\label{sol2a}
A(t)&=& A_0(t) +\alpha' A_1(t)+\alpha'^2 A_2(t)+ O(\alpha'^3),\br
B(t)&=& B_0(t) +\alpha' B_1(t)+\alpha'^2 B_2(t)+ O(\alpha'^3),\br
\Phi(t) &=& \Phi_0(t) + \alpha' \Phi_1(t) + \alpha'^2 \Phi_2(t) + O(\alpha'^3)
\eea 
where we  take $A_0, B_0, \Phi_0, A_1, B_1, \Phi_1$ as the known lower 
order solutions
worked out in the previous section. As an illustration we consider
the case of the background $(i)$, but similar computation can be done for
the background $(ii)$ also. Substituting these ansatze in the 
equations \eqn{Aeqn}, \eqn{Beqn} and \eqn{Feqn}
and collecting the coefficients of $\alpha'^2$ terms and equating them to 
zero, we find three equations involving $A_2(t)$, $B_2(t)$ and $\Phi_2(t)$.
Solving those equations we obtain the second order corrections. The complete
corrected solutions upto order $\alpha'^2$ are
\bea \label{sol4}
&&A(t)=t^{5\over 9}\left(1-{775\over 13122} {\alpha' \over t^2}-{2125975\over 
172186884}{\alpha'^2\over t^4}\right)+O(\alpha'^3),\br
&& B(t)=t^{-{1/9}}\left(1+{10\over 6561}{\alpha' \over t^2} + {89975\over 
172186884}{\alpha'^2\over t^4}\right)+O(\alpha'^3),\br
&& \Phi(t) = g_s^{-2}\left(1 + {115\over 1458}{\alpha' \over t^2}+
{51625\over 2125764}{\alpha'^2\over t^4}\right) + O(\alpha'^3)
\eea
Similarly the curvature and the Gauss-Bonnet term become
\bea
&& R={1\over  t^2}\left(-{230\over 243}{\alpha' \over t^2}- {482075\over 
531441}{\alpha'^2 \over 
t^4}+O(\alpha'^3)\right)\br
&& R_{GB}^2={1\over  t^4}\left({1840\over 729} + {1320800\over
531441}{\alpha' \over
t^2}+O(\alpha'^2)\right)\br
&& R_{MN} R^{MN} = {68350\over 177147} {\alpha'\over t^6} + O(\alpha'^2)\br
&& R_{MNPQ} R^{MNPQ} = {1\over t^4}\left({1840\over 729}
+ {1320800\over
531441}{\alpha' \over
t^2}+O(\alpha'^2)\right) 
\eea
We can see that this expansion could be arranged as powers of ${\alpha' 
\over t^2}$. Also the coefficients do not change signs as we go up higher 
in the order.    

Let us also note that the complete on-shell action upto second order 
corrections can 
now be expressed as 
\be
S\sim \int dt {1\over t} \left[ \sum_{n=0}^\infty a_n \left({\alpha' \over 
t^2}\right)^n\right]    
\ee
where some of the calculated coefficients are $a_0=0,~a_1=-{460\over 
729},~ a_2=-{287075\over 531441}$. The series appears to be convergent.

\subsection{Non-constant lowest order dilaton}

Having studied the Kasner solutions with the lowest order dilaton taken 
to be constant in the previous subsections we 
would like to evalaute the higher derivative corrections to
the time-dependent 
heterotic S$p$-brane solutions with non-constant lowest order dilaton field. 
The complete 
action upto $\alpha'$ order has been given in \eqn{action}.  
Using the same metric ansatz in the action we get the same equations of
motion \eqn{Aeqn}, \eqn{Beqn} and \eqn{Feqn}. However since in this case
the lowest order dilaton is non-constant, the zeroth order equation in 
$\alpha'$ will involve $\Phi_0$ and its derivatives unlike the similar
equations \eqn{cdzero} in the previous case. Substituting \eqn{sol2}
in \eqn{Aeqn}, \eqn{Beqn} and \eqn{Feqn} we get the zeroth order equations
in the form,
\bea\label{zeroth}
&&30 \Phi_0^2 A_0^2 {\dot B}_0^2+12 \Phi_0 B_0 A_0
\left(2 \Phi_0 {\dot B}_0 {\dot A}_0+A_0 \left({\dot\Phi}_0 {\dot B}_0+
\Phi_0 {\ddot B}_0\right)\right)\br
&&+B_0^2
\left(2 \Phi_0^2 {\dot A}_0^2-A_0^2 \left({\dot\Phi}_0^2-2 \Phi_0 {\ddot\Phi}_0
\right)+4 \Phi_0 A_0
\left({\dot\Phi}_0 {\dot A}_0+\Phi_0 {\ddot A}_0\right)\right) = 0\br\br
&&20 \Phi_0^2 A_0^2 {\dot B}_0^2+10 \Phi_0
B_0 A_0 \left(3 \Phi_0 {\dot B}_0 {\dot A}_0+A_0 \left({\dot\Phi}_0 {\dot B}_0
+\Phi_0
{\ddot B}_0\right)\right)\br
&&+B_0^2
\left(6 \Phi_0^2 {\dot A}_0^2-A_0^2 \left({\dot\Phi}_0^2-2 \Phi_0 {\ddot\Phi}_0
\right)
+6 \Phi_0 A_0 \left({\dot\Phi}_0 {\dot A}_0+\Phi_0 {\ddot A}_0
\right)\right) = 0\br\br
&&30 \Phi_0^2 A_0^2 {\dot B}_0^2+12 \Phi_0 B_0
A_0 \left(3 \Phi_0 {\dot B}_0 {\dot A}_0+A_0 \left({\dot\Phi}_0 {\dot B}_0
+\Phi_0 {\ddot B}_0
\right)\right)\br
&&+B_0^2
\left(6 \Phi_0^2 {\dot A}_0^2-A_0^2 \left({\dot\Phi}_0^2-2 \Phi_0 
{\ddot\Phi}_0\right)
+6 \Phi_0 A_0 \left({\dot\Phi}_0 {\dot A}_0+{\dot\Phi}_0 {\ddot A}_0
\right)\right) = 0
\eea
Now it can be easily checked that both the solutions given in 
eqs.\eqn{string} \eqn{2ndstring} indeed solve the above equations 
\eqn{zeroth}
as they should. Substituting the first solution $(i')$ of \eqn{2ndstring}
in \eqn{zeroth} we get three equations at the order $\alpha'$ involving 
$A_1(t)$, $B_1(t)$
and $\Phi_1(t)$ as follows,
\bea\label{first}
&&-\frac{6560}{729 t^{8/3}}+\frac{88 \Phi_1}
{t^{14/3}}+\frac{64 B_1}{81 t^{1/3}}+\frac{64 A_1}{243
t^{1/3}}-\frac{40 {\dot\Phi}_1}{t^{11/3}}+\frac{320}{27} t^{2/3} 
{\dot B}_1+\frac{320}{81} t^{2/3} {\dot A}_1\br
&&+\frac{6 
{\ddot\Phi}_1}{t^{8/3}}+\frac{64}{9}
t^{5/3} {\ddot B}_1+\frac{64}{27} t^{5/3} 
{\ddot A}_1= 0\br
&&\frac{13120}{729 t^{8/3}}+\frac{176 
\Phi_1}{t^{14/3}}+\frac{320 B_1}{243 t^{1/3}}
+\frac{64 A_1}{81
t^{1/3}}-\frac{80 {\dot\Phi}_1}{t^{11/3}}+\frac{1600}{81} t^{2/3} 
{\dot B}_1+\frac{320}{27} t^{2/3} {\dot A}_1\br
&&+\frac{12 
{\ddot\Phi}_1}{t^{8/3}}+\frac{320}{27}
t^{5/3} {\ddot B}_1+\frac{64}{9} t^{5/3} {\ddot A}_1
= 0\br
&&-\frac{1070}{9 t^7}+\frac{162 
\Phi_1}{t^9}-\frac{567 {\dot\Phi}_1}{8 t^8}+\frac{16 
{\dot B}_1}{t^{11/3}}+\frac{8
{\dot A}_1}{t^{11/3}}+\frac{81 {\ddot \Phi}_1}{8 t^7}+
\frac{12 {\ddot B}_1}{t^{8/3}}+\frac{6 {\ddot A}_1}{t^{8/3}}
= 0
\eea
The equations \eqn{first} can be solved and we find,
\be
A_1(t) = B_1(t) = \frac{755}{54} t^{-\frac{7}{3}} \quad \Phi_1(t) = 
-\left(\frac{2}{3}\right)^4 \frac{2375}{18} t^2
\ee
So, the complete solution upto first order in $\alpha'$ is
\be A(t) = B(t) = t^{-\frac{1}{3}}\left(1+\frac{755}{54}\frac{\alpha'}{t^2}
\right), \qquad \Phi(t) = \left(\frac{2}{3}t\right)^4 \left( 1 - 
\frac{2375}{18}\frac{\alpha'}{t^2}\right)
\ee
It is interesting to note that the $\alpha'$ corrections do not break the 
$SO(9)$ invariance of the original solution given in $(i')$ of 
\eqn{2ndstring}. The other covariant quantities of 
interest are
\bea
&&{\rm Det}(-g)= \frac{1}{t^6} \left(1+{755\over 3}{ \alpha'\over t^2}\right)
+O(\alpha'^2)\br
&& R={1\over t^2}\left(16+{28690\over 9}{\alpha'\over t^2}\right) +
O(\alpha'^2)\br   
&& R_{MNPQ}R^{MNPQ}={80\over 9 t^4}+O(\alpha')\br   
&& R_{MN}R^{MN}= {32\over  t^4}+O(\alpha')\br   
&& R^2_{GB}={1232\over 9}{1\over t^4}+O(\alpha')\ .
\eea
From the determinant of the metric we determine that there is a cut-off 
time $t=t_g$ below which the calculations cannot be trusted, means 
perturbative approximation will break down. It is given by
\be
(t_g)^2={755\over 3}{ \alpha'}
\ee
For the time range $t>t_g$ all the curvature expressions stay finite. For 
large time all the $\alpha'$ corrections to the solution become 
negligible and the asymptotic generalized Kasner background emerges.

By using the similar technique we can also obtain the corrections of the 
second solution $(ii')$ of \eqn{2ndstring}. We here give the complete solution
as,
\bea
&& A(t) = t^{-\frac{1}{\sqrt {3}}}\left\{1 +\left(\frac{111+80
\sqrt{3}}{72}\right) \frac{\alpha'}{t^2}\right\}\br
&& B(t) = 1+\left(\frac{14+9\sqrt{3}}{8}\right)
\frac{\alpha'}{t^2}\br
&& \Phi(t) = t^{\left(1+\frac{1}{\sqrt{3}}\right)}\left\{1-
\left(\frac{365+236\sqrt{3}}{24}\right)\frac{\alpha'}{t^2}\right\}
\eea
The other quantities of interest are
\bea
&&{\rm Det}(-g)= t^{-2\sqrt{3}} \left\{1+\left({121(\sqrt{3}+2)
\over 4\sqrt{3}}\right)
{ \alpha'\over t^2}\right\}
+O(\alpha'^2)\br
&& R={1\over t^2}\left(4+2\sqrt{3}\right) + \left({1895 + 1126\sqrt{3} 
\over 6}\right)
{\alpha' \over t^4}+ 
O(\alpha'^2)\br
&& R_{MNPQ}R^{MNPQ}={1\over 3 t^4}\left(20+8\sqrt{3}\right)+O(\alpha')\br
&& R_{MN}R^{MN}= {1\over  t^4} \left(8+4\sqrt{3}\right)+O(\alpha')\br
&& R^2_{GB}={8\over 3t^4}\left(1+\sqrt{3}\right)+O(\alpha')
\eea
We again find from the determinant of the metric that the cut-off time
in this case is given by
\be
t_g^2 = \frac{121(\sqrt{3}+2)}{4\sqrt{3}} \alpha'
\ee
below which the perturbative approximation will break down. However,
for $t>t_g$ the curvature invariants remain finite. For large time
we again recover the generalized Kasner form of the background.  

As indicated in the case of cosmologies with the constant lowest order dilaton 
studied in the previous
subsections, here also 
the more higher order $\alpha'$ corrections can be computed. The 
general comments about the
spontaneous compactification as well as the improvement on the deceleration
rate remain more or less similar in this case also.  

\section{Summary}

In this paper we have studied the $\alpha'$ corrections to a class of time
dependent solutions called space-like or S-branes in string theory. S-branes
generically have a space-like singularity at $t=0$. The near `horizon' or
near $t=0$ limit of these S-branes have the structure of the generalized Kasner
geometry which are the global solutions of string theory with singularity
at $t=0$. Since for these solutions curvature blows up at $t=0$, 
we have included the
higher order curvature terms or the $\alpha'$ correction terms to see their
effects on the geometry. We have considered the heterotic string theory
S$p$-brane solutions since for this case the exact correction terms
upto $\alpha'^2$ order
are known. This is in contrary to the type II string theory, where the
first non-trivial correction comes at the order of $\alpha'^3$ and the
calculation becomes quite involved. We have used both the constant lowest
order dilaton
i.e. the usual Kasner like solutions and the non-constant dilaton i.e. 
the generalized
Kasner like solutions. In both cases we found the corrected geometries
when the $\alpha'$ corrections are included
in the action. The interesting thing is that by obtaining the corrections, 
we are able to determine the time  $t=t_g$ when the curvature starts 
becoming large for the cases we have studied. For the time $t>t_g$ 
the spacetime is finite and supergravity is a valid approximation.
In the regime $t\le t_g$ the spacetime curvature becomes higher and string 
theory is the only valid description there. 
Although, the  S$p$-brane backgrounds we 
have studied exist only for $t > 0$, it will be nevertheless interesting 
to 
study those solutions which are explicitly time-reversal symmetric.      

While studying spontaneous compactification we find that the 
Kasner cosmologies 
are generally decelerating as usual, but the deceleration is faster initially 
and could be made close to value in the radiation dominated phase
of our universe if the higher order corrections are included. 

\section*{Note added:}

After submission of this paper to the Archive we received a very useful
correspondence from Qasem Exirifard in which
he drew our attention to some of the issues we were not aware of.
This has helped us, we hope, to improve our paper.


\begin{thebibliography}{99}

\bibitem{Roy:2006du}
  S.~Roy and H.~Singh,
  ``Space-like branes, accelerating cosmologies and the near 'horizon' limit,''
  JHEP {\bf 0608}, 024 (2006)
  [arXiv:hep-th/0606041].

\bibitem{Gutperle:2002ai}
  M.~Gutperle and A.~Strominger,
  ``Spacelike branes,''
  JHEP {\bf 0204}, 018 (2002)
  [arXiv:hep-th/0202210].

\bibitem{Chen:2002yq}
  C.~M.~Chen, D.~V.~Gal'tsov and M.~Gutperle,
  ``S-brane solutions in supergravity theories,''
  Phys.\ Rev.\  D {\bf 66}, 024043 (2002)
  [arXiv:hep-th/0204071].

\bibitem{Kruczenski:2002ap}
  M.~Kruczenski, R.~C.~Myers and A.~W.~Peet,
  ``Supergravity S-branes,''
  JHEP {\bf 0205}, 039 (2002)
  [arXiv:hep-th/0204144].

\bibitem{Bhattacharya:2003sh}
  S.~Bhattacharya and S.~Roy,
  ``Time dependent supergravity solutions in arbitrary dimensions,''
  JHEP {\bf 0312}, 015 (2003)
  [arXiv:hep-th/0309202].

\bibitem{Roy:2002ik}
  S.~Roy,
  ``On supergravity solutions of space-like Dp-branes,''
  JHEP {\bf 0208}, 025 (2002)
  [arXiv:hep-th/0205198].

\bibitem{Sen:2002nu}
  A.~Sen,
  ``Rolling tachyon,''
  JHEP {\bf 0204}, 048 (2002)
  [arXiv:hep-th/0203211].

\bibitem{Sen:2002in}
  A.~Sen,
  ``Tachyon matter,''
  JHEP {\bf 0207}, 065 (2002)
  [arXiv:hep-th/0203265].

\bibitem{Sen:2002an}
  A.~Sen,
  ``Field theory of tachyon matter,''
  Mod.\ Phys.\ Lett.\  A {\bf 17}, 1797 (2002)
  [arXiv:hep-th/0204143].

\bibitem{Townsend:2003fx}
  P.~K.~Townsend and M.~N.~R.~Wohlfarth,
  ``Accelerating cosmologies from compactification,''
  Phys.\ Rev.\ Lett.\  {\bf 91}, 061302 (2003)
  [arXiv:hep-th/0303097].

\bibitem{Ohta:2003pu}
  N.~Ohta,
  ``Accelerating cosmologies from S-branes,''
  Phys.\ Rev.\ Lett.\  {\bf 91}, 061303 (2003)
  [arXiv:hep-th/0303238].

\bibitem{Roy:2003nd}
  S.~Roy,
  ``Accelerating cosmologies from M/string theory compactifications,''
  Phys.\ Lett.\  B {\bf 567}, 322 (2003)
  [arXiv:hep-th/0304084].


\bibitem{Ohta:2003ie}
  N.~Ohta,
  ``A study of accelerating cosmologies from superstring / M theories,''
  Prog.\ Theor.\ Phys.\  {\bf 110}, 269 (2003)
  [arXiv:hep-th/0304172].

\bibitem{Maeda:2004hu}
  K.~i.~Maeda and N.~Ohta,
  ``Inflation from superstring / M theory compactification with higher  order
  corrections. I,''
  Phys.\ Rev.\  D {\bf 71}, 063520 (2005)
  [arXiv:hep-th/0411093].

\bibitem{Akune:2006dg}
  K.~Akune, K.~i.~Maeda and N.~Ohta,
  ``Inflation from superstring / M-theory compactification with higher order
  corrections. II: Case of quartic Weyl terms,''
  Phys.\ Rev.\  D {\bf 73}, 103506 (2006)
  [arXiv:hep-th/0602242].

\bibitem{Niz:2007gq}
  G.~Niz and N.~Turok,
  ``Stringy corrections to a time-dependent background solution of string 
and
  M-theory,''
  Phys.\ Rev.\  D {\bf 75}, 126004 (2007)
  [arXiv:0704.1727 [hep-th]].


\bibitem{Dabholkar:2004yr}
  A.~Dabholkar,
  ``Exact counting of black hole microstates,''
  Phys.\ Rev.\ Lett.\  {\bf 94}, 241301 (2005)
  [arXiv:hep-th/0409148].

\bibitem{Dabholkar:2004dq}
  A.~Dabholkar, R.~Kallosh and A.~Maloney,
  ``A stringy cloak for a classical singularity,''
  JHEP {\bf 0412}, 059 (2004)
  [arXiv:hep-th/0410076].

\bibitem{Sen:2004dp}
  A.~Sen,
  ``How does a fundamental string stretch its horizon?,''
  JHEP {\bf 0505}, 059 (2005)
  [arXiv:hep-th/0411255].

\bibitem{Sen:2005pu}
  A.~Sen,
  ``Black holes, elementary strings and holomorphic anomaly,''
  JHEP {\bf 0507}, 063 (2005)
  [arXiv:hep-th/0502126].

\bibitem{Sen:2005wa}
  A.~Sen,
  ``Black hole entropy function and the attractor mechanism in higher
  derivative gravity,''
  JHEP {\bf 0509}, 038 (2005)
  [arXiv:hep-th/0506177].

\bibitem{Dabholkar:2006tb}
  A.~Dabholkar, A.~Sen and S.~P.~Trivedi,
  ``Black hole microstates and attractor without supersymmetry,''
  JHEP {\bf 0701}, 096 (2007)
  [arXiv:hep-th/0611143].

\bibitem{Astefanesei:2006sy}
  D.~Astefanesei, K.~Goldstein and S.~Mahapatra,
  ``Moduli and (un)attractor black hole thermodynamics,''
  arXiv:hep-th/0611140.

\bibitem{Das:2006dz}
  S.~R.~Das, J.~Michelson, K.~Narayan and S.~P.~Trivedi,
  ``Time dependent cosmologies and their duals,''
  Phys.\ Rev.\  D {\bf 74}, 026002 (2006)
  [arXiv:hep-th/0602107].

\bibitem{Exirifard:2004ey}
  G.~Exirifard and M.~O'Loughlin,
  ``Two and three loop alpha' corrections to T-duality: Kasner and
  Schwarzschild,''
  JHEP {\bf 0412}, 023 (2004)
  [arXiv:hep-th/0408200].

\bibitem{Freeman:1986zh}
  M.~D.~Freeman, C.~N.~Pope, M.~F.~Sohnius and K.~S.~Stelle,
  ``Higher order sigma model counterterms and the effective action for
  superstrings,''
  Phys.\ Lett.\  B {\bf 178}, 199 (1986);
  M.~T.~Grisaru, A.~E.~M.~van de Ven and D.~Zanon,
  ``Four loop beta function for the N=1 and N=2 supersymmetric nonlinear sigma
  model in two-dimensions,''
  Phys.\ Lett.\  B {\bf 173}, 423 (1986);
  Q.~H.~Park and D.~Zanon,
  ``More on sigma model beta functions and low-energy effective actions,''
  Phys.\ Rev.\  D {\bf 35}, 4038 (1987);
  I.~Jack, D.~R.~T.~Jones and D.~A.~Ross,
  ``On the relationship between string low-energy effective actions and O
  (alpha-prime**3) sigma model beta functions,''
  Nucl.\ Phys.\  B {\bf 307}, 130 (1988).

\bibitem{Gross:1986mw}
  D.~J.~Gross and J.~H.~Sloan,
  ``The quartic effective action for the heterotic string,''
  Nucl.\ Phys.\  B {\bf 291}, 41 (1987).

\bibitem{Hull:1987yi}
  C.~M.~Hull and P.~K.~Townsend,
  ``String effective actions from sigma model conformal anomalies,''
  Nucl.\ Phys.\  B {\bf 301}, 197 (1988).

\bibitem{Callan:1988hs}
  C.~G.~Callan, R.~C.~Myers and M.~J.~Perry,
  ``Black Holes in String Theory,''
  Nucl.\ Phys.\  B {\bf 311}, 673 (1989).

\end{thebibliography}
\end{document}